# Why Observable Space Is Solely Three Dimensional

Mario Rabinowitz

Armor Research, 715 Lakemead Way
Redwood City, CA 94062-3922, USA

**Abstract**

Quantum (and classical) binding energy considerations in n-dimensional space indicate that atoms (and planets) can only exist in three-dimensional space. This is why observable space is solely 3-dimensional. Both a novel Virial theorem analysis, and detailed classical and quantum energy calculations for 3-space circular and elliptical orbits indicate that they have no orbital binding energy in greater than 3-space. The same energy equation also excludes the possibility of atom-like bodies in strictly 1 and 2-dimensions. A prediction is made that in the search for deviations from $r^{-2}$ of the gravitational force at sub-millimeter distances such a deviation must occur at $< \sim 10^{-10}$m (or $< \sim 10^{-12}$ m considering muoniom), since atoms would disintegrate if the curled up dimensions of string theory were larger than this. Callender asserts that the often-repeated claim in previous work that stable orbits are possible in only three dimensions is not even remotely established. The binding energy analysis herein avoids the pitfalls that Callender points out, as it circumvents stability issues. An uncanny quantum surprise is present in very high dimensions.

**Keywords:** Space dimensionality, No binding energy in higher dimensions, Atoms, Absence of orbits, Virial theorem, Higher space, Quantum surprise

## 1 Introduction

Atoms (and planets) are bound by a $r^{-1}$ potential in 3-space. Atoms (as we know them) are necessary for the existence of observers of our universe. The $r^{-1}$ 3-space potential will be generalized to n-dimensional Euclidean space to determine the effect of other dimensions. It will be shown that observable space is not more than three-dimensional because atoms (and planets) cannot exist in higher dimensional space. Both a novel Virial theorem analysis, and detailed classical and quantum energy calculations indicate that there is no orbital binding energy in greater than 3-space.

Callender in his penetrating review and critique [3] concludes that n-space arguments based on orbital stability of either planets or atomic orbital electrons fall short of the mark in asserting that space must be three-dimensional. He asserts that the often repeated claim that stable orbits are possible in only three dimensions is not even remotely established. The binding energy analysis herein circumvents Callender's criticisms.

If the sum of kinetic energy + potential energy $\equiv E_n$ is negative for an orbiting body, the body is in a bound orbit, and $E_n$ is the binding energy. If $E_n \geq 0$, the body is not bound. Although intelligent observers, could in principle, live on non-closed gravitational orbits, life as we know it could not exist without atoms.

## 2  Background

In 1746 Immanuel Kant wrote, "substances in the existing world…have essential forces of such a kind that they propagate their effects in union with each other according to the inverse-square relations of the distances, …[and] that the whole to which this gives rise has, by virtue of this law, the property of being three-dimensional". [3] His reasoning could also be inverted that 3-dimensional space implies the inverse-square law of force, as the argument goes both ways.

The argument in this paper is of a different kind. It will be shown that atoms (and planetary systems) as we know them in 3-space (with a $r^{-2}$ force holding them together and a $r^{-1}$ potential), cannot exist in higher (or lower) dimensional space. Therefore if higher dimensional space exists, we could not exist in it, nor could any other entities made of atoms. This is why we can only directly observe 3-dimensional space, and a 4-dimensional space-time manifold. The effect of non-Euclidean general relativity will be discussed in Sec. 6.

In the modern era, one of the earliest percipient physics papers on why space has three dimensions was written in 1917 by Ehrenfest [5]. His conclusion that stable orbits are possible only in three dimensions is critiqued by Callender [3].

As a small part of papers in 2001 and 2007, Rabinowitz [7, 8] introduced n-dimensional binding energy analysis of quantized tiny gravitational orbits; and indicated that the analysis also applies to ordinary atoms. This analysis showed that such orbits could not be bound energetically for n ≥ 4, and also excluded n = 1, 2. The present paper goes into greater depth as well as considering questions not covered in those papers.

## 3 Novel Virial Theorem Analysis

The Coulomb electric field of a charge $+ze$ acting on a charge $-e$ in 3-space produces a force (mks units) between the charges:

$$F_3 = \frac{-ze^2}{4\pi\varepsilon_0 r_3^2}, \tag{1}$$

with potential energy $V_3 = \frac{-ze^2}{4\pi\varepsilon_0 r_3}$, where $\varepsilon_0$ is the permittivity of free 3-space, and $r_3$ is the separation between $ze$ and $-e$. In n-space, $\varepsilon_{0n}$ is the free space permittivity, and also incorporates the dimensionality.

The 3-dimensional Coulomb force can be generalized to n-dimensional Euclidean space, by means of Gauss' law using the generalized area $A_{n-1}$ as derived in [3].

$$\oint \frac{F_n}{-e} \cdot dA_{n-1} = \frac{ze}{\varepsilon_0}, \tag{2}$$

$$\Rightarrow F_n = \frac{-ze^2}{\varepsilon_0 A_{n-1}} = \frac{-ze^2 \Gamma(n/2)}{2\varepsilon_0 \pi^{n/2} r_n^{n-1}}, \text{ where} \tag{3}$$

$$A_{n-1} = \frac{2\pi^{n/2} r_n^{n-1}}{\Gamma(n/2)}, \tag{4}$$

and the Gamma function $\Gamma(n) \equiv \int_0^\infty t^{n-1} e^{-t} dt$ for all n (integer and noninteger). When n is an integer, $\Gamma(n) = (n-1)!$.

Similarly for Newton's 3-dimensional gravitational equation (the force between two masses M and m separated by a distance $r_3$ where $G_3$ is the 3-space universal gravitational constant)

$$F_3 = \frac{-G_3 Mm}{r_3^2}, \tag{5}$$

$$\oint \frac{F_n}{m} \cdot dA_{n-1} = -4\pi G_n M, \tag{6}$$

$$\Rightarrow F_n = \frac{-4\pi G_n Mm}{A_{n-1}} = \frac{-4\pi G_n Mm \Gamma(n/2)}{2\pi^{n/2} r_n^{n-1}}. \tag{7}$$

The electric and gravitational forces can be written as one equation in terms of a generalized representation by introducing a source term $K > 0$:

$$F_n = \frac{-K}{A_{n-1}} = \frac{-K\Gamma(n/2)}{2\pi^{n/2}r_n^{n-1}}, \tag{8}$$

For the Electric Field: $K = K_E = ze^2/\varepsilon_0$ in mks units.

For the gravitational field $K = K_G = 4\pi G_n Mm$ where the gravitational constant $G_n$ is dimensional and model dependent for $n > 3$; M is the mass that produces the gravitational field, and m << M is the mass orbiting M. When m ~ M, the mass m can be replaced by the reduced mass $\mu = \frac{mM}{m+M}$ in the equations of motion, accounting for motion of both m and M about their common center of mass. In either case m << M or m ~ M, central force motion implies the motion is in a plane in n-space because there is no torque.

As we have seen, the power of r in the force equation $F_n$ is $\propto r_n^{-n+1}$, where $n$ is the number of spatial dimensions in the space-time manifold of $(n+1)$ dimensions. We will only be concerned with a single particle moving under a central force $F_n = -\nabla V_n$ so the potential energy $V_n = C_n r_n^{-n+2}$, where $C_n$ depends on both the interaction and the dimensionality. In this form, the exponent of $r_n$ is directly tied into the n-dimensionality of space. Thus in general, we can write the Virial theorem as

$$\langle V_n \rangle = 2\langle T_n \rangle /(-n+2) \tag{9}$$

$$\Rightarrow \langle E_n \rangle = \langle T_n \rangle + \langle V_n \rangle = \left(\frac{n-4}{n-2}\right)\langle T_n \rangle, \tag{10}$$

where $\langle T_n \rangle$ is the time average value of the kinetic energy of the particle in n-space. The Virial theorem is based on the concept of periodic motion. It was derived in 1870 by Rudolph Clausius with great generality [4]. This makes it ideal for the analysis of binding energy $E_n$ in n-space. If the motion is assumed to be periodic and a contradiction is obtained (such as when $E_n \geq 0$) then the motion is not periodic because it is unbounded. This kind of dichotomy is also used later in this paper.

From Eq. (10), we can conclude that the time average energy $\langle E_n \rangle \geq 0$ for $n \geq 4$. This implies that for a time average over an orbit, it is not energetically bound because on the average there is no binding energy. In the case $\langle E_4 \rangle = 0$, a small energy increasing perturbation can destroy the orbit; so the orbit is essentially unbound. So orbiting entities that are bound by a $r^{-1}$ potential in 3-space, are simply unbound in more than 3 dimensions. Therefore, planetary systems and atoms cannot exist in higher dimensions than 3.

The advantage of the time average Virial theorem approach is that it is completely general for all orbits. In 3-space where $r^{-1}$ bound orbits are circular or elliptical, the Virial theorem handles them both with ease. However to some, time averages may not seem sufficiently compelling for the conclusion reached. So we will next look at actual values of the variables rather than time averages.

## 4 General Classical Analysis in Euclidean Space

If a body bound in a circular or elliptical orbit in 3-space were to enter higher dimensional space, what would happen to its orbit? For an elliptical orbit, the centripetal and centrifugal forces are not equal at any point of the orbit; whereas they are equal everywhere on a circular orbit. For convenience we will start with a circular orbit. Then a connection between circular and elliptic motion will be made. For a body in circular orbit in a central force field, we can easily obtain the rotational kinetic energy $T_n$ from the centripetal force equation (8) in n-space:

$$F_n = \frac{-K\Gamma(n/2)}{2\pi^{n/2}r^{n-1}} = \frac{-mv_n^2}{r_n} \tag{11}$$

$$\Rightarrow \tfrac{1}{2}mv_n^2 \equiv T_n = \tfrac{1}{2}F_n r_n = \frac{K\Gamma(n/2)}{4\pi^{n/2}r_n^{n-1}} r_n = \frac{K\Gamma(n/2)}{4\pi^{n/2}r_n^{n-2}}, \tag{12}$$

where $v_n$ is the magnitude of the rotational velocity. The potential energy is:

$$V_n = -\int F_n \cdot dr_n = \frac{-K\Gamma(n/2)}{2(n-2)\pi^{n/2}r_n^{n-2}} = \frac{-2T_n}{(n-2)}. \tag{13}$$

This implies that the total energy $E_n$ associated with a circular orbit is:

$$E_n = T_n + V_n = T_n + \frac{-2T_n}{(n-2)} = \frac{(n-4)}{(n-2)}T_n. \tag{14}$$

In 3-space, the energy of an elliptic orbit is equal to the energy of a circular orbit when the semi-major axis $a_{3e}$ of the elliptic orbit is equal to the radius $r_{3c}$ of the circular orbit [1]:

$$E_3 = \frac{-K}{2a_{3e}} = \frac{-K}{2r_{3c}}. \tag{15}$$

There is an $a_{3e}$ for every $r_{3c}$. Thus in 3-space, Eq. (14) holds for all bound orbits (circular and elliptical) classically and will also found to be valid for circular and elliptical orbits in the Bohr quantum equations for all quantum numbers j (cf. Sec. 5). When we find that the total energy $E_n = (n-4)T_n/(n-2) \geq 0$, where $T_n$ is the kinetic energy, we may conclude that formerly bound orbits cannot be bound energetically when introduced into the given n-space. Eq. (14) is an extension of Eq. (10) from time averages to actual values. $E_n \geq 0$ occurs for $n \geq 4$. In 4-space, $E_4 = 0$, because the kinetic energy and potential energy cancel each other. As will be shown by Eq. (24), the quantum orbit radius $r_4$ is infinite, implying that the 4-space orbit is unbound.

In the gravitational case (substituting for $T_n$ from Eq. (12) into Eq. (14) with $K = 4\pi G_n Mm$) we have

$$E_n = \frac{(n-4)}{(n-2)}\left[\frac{K\Gamma\left(\frac{n}{2}\right)}{4\pi^{n/2}r_n^{n-2}}\right] = \frac{(n-4)}{(n-2)}\left[\frac{4\pi G_n Mm\Gamma\left(\frac{n}{2}\right)}{4\pi^{n/2}r_n^{n-2}}\right] \geq 0 \text{ for } n \geq 4. \tag{16}$$

## 5  n-Dimensional Bohr Circular and Elliptical Quantized Atomic Orbits

Next we will see that electrons cannot be energetically bound in atoms for $n \geq 4$. For an electrostatic elliptical orbit in 3 space, the total energy is

$$E_3 = \frac{-ze^2}{2a_{3e}(4\pi\varepsilon_0)} = \frac{-ze^2}{2r_{3c}(4\pi\varepsilon_0)}, \tag{17}$$

where $a_{3e}$ is the major semi-axis of the ellipse, and $r_{3c}$ is the corresponding circular radius, which in this quantum case is the Bohr orbit. Bohr-Sommerfeld quantization is not necessary for the elliptical orbit, as Born [1] derives both $a_{3e}$ and $r_{3c}$ using just Bohr quantization

$$r_3 \equiv r_{3c} = a_{3e} = \frac{(4\pi\varepsilon_0)\hbar^2 j^2}{mze^2}, \tag{18}$$

where $j = 1, 2, 3, \ldots$ is the quantum number for the $j$ th energy state. It is relevant to note that the elliptical orbits in Bohr-Sommerfeld quantization have the same energy levels as the circular orbits of Bohr quantization.

Empirically, Bohr quantum theory gives highly accurate energy solutions for the atom. This is remarkable since in many other respects the Bohr quantum theory is not only less accurate, it is less consistent than modern quantum theory [6]. This happens in those cases where it predicts that an electron goes straight through the atomic nucleus, and in its artificial introduction of zero-point energy. Nevertheless its use in the context of energy solutions is objectively justifiable.

It is noteworthy that aside from small corrections, the Bohr atom gives atomic energies with good accuracy [8]. For an additional example it yields the Balmer term in atomic spectra with the correct coefficient. Since the argument herein depends only on energies, we may proceed with our n-dimensional analysis using Bohr quantization for circular orbits. Substituting Eq. (15) into (14) accurately gives the quantized total electrostatic energy for either a circular or an elliptical orbit in 3-space:

$$E_3 = \frac{-ze^2}{2r_3(4\pi\varepsilon_0)} = \frac{-z^2 e^4 m}{2(4\pi\varepsilon_0)^2 \hbar^2 j^2}. \tag{19}$$

For an electrostatically bound atom, we equate the electrostatic and centripetal forces in n-space:

$$F_n = \frac{-ze^2}{\varepsilon_0 A_{n-1}} = \frac{-ze^2 \Gamma(n/2)}{\varepsilon_0 2\pi^{n/2} r_n^{n-1}} = \frac{-mv_n^2}{r_n}, \tag{20}$$

with $A_{n-1}$ given by Eq. (4). Solving Eq. (20) for the rotational velocity magnitude in a circular orbit

$$v_n = \left[\frac{ze^2 \Gamma\left(\frac{n}{2}\right)}{m\varepsilon_0 2\pi^{n/2} r_n^{n-2}}\right]^{1/2} = \left[\frac{ze^2 \Gamma\left(\frac{n}{2}\right)}{2m\varepsilon_0 \pi^{n/2}}\right]^{1/2} r_n^{\frac{(2-n)}{2}}. \tag{21}$$

Quantization of angular momentum:

$$mv_n r_n = j\hbar. \tag{22}$$

Substituting Eq. (21) into (22)

$$m\left[\frac{ze^2 \Gamma\left(\frac{n}{2}\right)}{2m\varepsilon_0 \pi^{n/2}}\right]^{1/2} r_n^{\frac{(2-n)}{2}} r_n = j\hbar. \tag{23}$$

$$\Rightarrow r_n = j^2 \hbar^2 \left[\frac{2\varepsilon_0 \pi^{n/2}}{mze^2 \Gamma\left(\frac{n}{2}\right)}\right]^{\left(\frac{1}{4-n}\right)}. \tag{24}$$

Solving Eq. (24) for n = 3

$$r_3 = \frac{j^2 \hbar^2 2\varepsilon_0 \pi^{3/2}}{mze^2 \left[\frac{1}{2}\pi^{1/2}\right]} = \frac{j^2 \hbar^2 4\pi\varepsilon_0}{mze^2} = \frac{j^2 \hbar^2}{mz(e^2/4\pi\varepsilon_0)}. \tag{25}$$

This agrees with a direct 3-dimensional calculation.

To obtain the magnitude of the rotational velocity of the circular orbit, we substitute Eq. (24) into (21):

$$v_n = \left[\frac{ze^2 \Gamma\left(\frac{n}{2}\right)}{2m\varepsilon_0 \pi^{n/2}}\right]^{1/2} r^{\frac{(2-n)}{2}} = \left[\frac{ze^2 \Gamma\left(\frac{n}{2}\right)}{2m\varepsilon_0 \pi^{n/2}}\right]^{1/2} \left[j^2 \hbar^2 \left[\frac{2\varepsilon_0 \pi^{n/2}}{mze^2 \Gamma\left(\frac{n}{2}\right)}\right]^{\left(\frac{1}{4-n}\right)}\right]^{\frac{(2-n)}{2}}. \tag{26}$$

$$v_3 = \left[\frac{ze^2 \frac{1}{2}\pi^{1/2}}{2m\varepsilon_0 \pi^{3/2}}\right]^{1/2} \left[j^2 \hbar^2 \left[\frac{2\varepsilon_0 \pi^{3/2}}{mze^2 \frac{1}{2}\pi^{1/2}}\right]\right]^{\frac{(-1)}{2}} = \left[\frac{ze^2}{4\pi\varepsilon_0}\right]^{1/2} \left[\frac{1}{j\hbar}\right] \left[\frac{ze^2}{4\pi\varepsilon_0}\right]^{1/2} = \frac{1}{j\hbar}\left[\frac{ze^2}{4\pi\varepsilon_0}\right].$$

$\Rightarrow v_3 = \alpha c$ for j = 1, and z =1, where in mks units, the fine structure constant $\alpha \equiv e^2/4\pi\varepsilon_0 \hbar c$. In all units $\alpha \approx 1/137$.

In Eq. (14) we found quite generally that classically the total energy $E_n$ is:

$$E_n = T_n + V_n = \frac{(n-4)}{(n-2)} T_n, \tag{14}$$

where $T_n$ is the kinetic energy of an electron of charge $-e$ in orbit around a nucleus of charge $+ze$. Although it may be expected that the general result Eq. (14) also applies quantum mechanically, we need to ascertain possible quantum surprises. $E_n$ needs to be determined quantum mechanically. In order for this problem to be tractable in n-space we will deal with Bohr quantization which closely gives the same results for atomic energies as modern quantum mechanics as discussed above in this section.

Substituting for $T_n$ from Eq. (12) into Eq. (14) for the electrostatic case

$$E_n = \frac{(n-4)}{(n-2)} \frac{(ze^2/\varepsilon_0)\Gamma(\frac{n}{2})}{4\pi^{n/2} r_n^{n-2}}. \tag{27}$$

Substituting Eq. (24) for $r_n$ yields

$$E_n = \frac{(n-4)}{(n-2)} \frac{(ze^2/\varepsilon_0)\Gamma(\frac{n}{2})}{4\pi^{n/2}} (j^2\hbar^2)^{(2-n)} \left[\frac{2\varepsilon_0 \pi^{n/2}}{mze^2\Gamma(\frac{n}{2})}\right]^{\left(\frac{2-n}{4-n}\right)}$$

$$= \frac{(n-4)}{(n-2)} \left\{ \left[\frac{(ze^2/\varepsilon_0)\Gamma(\frac{n}{2})}{4\pi^{n/2}(j^2\hbar^2)^{(n-2)}}\right] \left[\frac{(ze^2/\varepsilon_0)m\Gamma(\frac{n}{2})}{2\pi^{n/2}}\right]^{\left(\frac{n-2}{4-n}\right)} \right\}. \tag{28}$$

The quantity in braces, $\{\ \} \geq 0$. For electrons to be bound we must have $E_n < 0$. Therefore for $n \geq 4$, all atomic orbitals (circular and elliptical) are unbound, i.e. atoms (as we know them) cannot exist in higher dimensional space. That is why observable space is not more than 3-dimensional. In 3-space, Eq. (28) gives the customary result:

$$E_3 = -\left[\frac{(ze^2/\varepsilon_0)(\frac{1}{2}\pi^{1/2})}{4\pi^{3/2}(j^2\hbar^2)}\right]\left[\frac{(ze^2/\varepsilon_0)m(\frac{1}{2}\pi^{1/2})}{2\pi^{3/2}}\right] = -\frac{(ze^2/4\pi\varepsilon_0)^2 m}{2j^2\hbar^2}. \tag{29}$$

This analysis showing that an atom with an electron of charge $e$ and mass $m$ orbiting the field of a nucleus of charge $ze$ cannot be bound energetically, is sufficiently general to include all atoms. This is because additional electrons would be at energy levels $\geq$ that of the single electron in the lowest energy state $j=1$, so that if the single electron is not energetically bound, neither will the additional electrons.

## 6 Discussion

It is noteworthy that the general Eq. (14) $E_n = (n-4)T_n/(n-2)$, not only establishes that atoms are not possible in higher dimensional space i.e. for n ≥ 4, but also makes a definitive statement about transforming 3-space atoms into solely 1 and 2-dimensional spaces (n = 1, 2). At n = 1, it states that $E_1 > 0$ implying that there are no bound particles in strictly 1-space. At n = 2, it states that $E_2 \to -\infty$, giving infinitely bound particles which is unphysical, thus excluding strictly 2-space. Furthermore in general relativity n = 1, 2 are excluded since it requires n ≥ 3 in order to have gravitation result from spatial curvature. Büchel [2] points out that life as we know it could not exist in 1-D or 2-D because of virtually impossible topological problems. Since the one formulation excludes atoms (as we know them) from existing in 1, 2, and n ≥ 4 dimensions, 3-space is the only dimension in which they can exist. So it was tempting to make the title of this paper: *Why Space Is Three Dimensional*. However such a title would be misleading. One reason is that all other dimensions except 3 are excluded is not the same thing as saying why space is 3-D. Inferable space is not excluded since it is not the same thing as observable space.

Shown below schematically are the limits for $E_n$ as $n \to \infty$. Note from Eq. (28) that quantum mechanically,

$$E_{QMn} \xrightarrow{n \to \infty} \frac{(n-4)}{(n-2)}\left[(ze^2/\varepsilon_0)m\Gamma(\tfrac{n}{2})\right]^{\left(\frac{n-1}{4-n}\right)} \xrightarrow{n \to \infty} \left[(ze^2/\varepsilon_0)m\Gamma(\tfrac{n}{2})\right]^{-1} \xrightarrow{n \to \infty} 0. \quad (30)$$

It is noteworthy that an uncanny quantum surprise is present. The source term (both the gravitational and electrostatic cases) becomes inverted, and $E_{QMn} \xrightarrow{n \to \infty} 0$. This remarkable inversion is worthy of further study.

Classically there is no inversion. From Eq. (27) for the electrostatic case for easy comparison,

$$E_{Cn} = \frac{(n-4)}{(n-2)}\left[\frac{(ze^2/\varepsilon_0)\Gamma(\tfrac{n}{2})}{4\pi^{n/2}r_n^{n-2}}\right] \xrightarrow{n \to \infty} \left[(ze^2/\varepsilon_0)\Gamma(\tfrac{n}{2})\right] \xrightarrow{n \to \infty} \infty. \quad (31)$$

From the Virial theorem Eq.(10), $\langle E_n \rangle = (n-4)\langle T_n \rangle/(n-2)$ involving time averages was readily obtained. Detailed classical and quantum calculations confirmed Eq. (14), $E_n = (n-4)T_n/(n-2)$ for actual rather than time averaged values. All these approaches lead to the same conclusion that there is no binding energy to hold a body in orbit for higher dimensional space: $E_n \geq 0$ for $n \geq 4$. A way to interpret the findings is that if a body bound in a circular or elliptical orbit in 3-space, were to enter higher dimensional space, because of flux divergence, it would become unbound and fly away from the central force. The methodology herein is to start with a potential and space (such as 3-space) that has bound orbits, then generalize this result to n-space to ascertain if these orbits are bound in other dimensions. Although traditionally treated the same mathematically, n dimensions in a higher dimensional space, are not the same as strictly n-space because flux can't be confined to just the given n dimensions in a higher space.

The generalization of Gauss' law to n-dimensions depends on a Euclidean isotropic space. The curvature of 4-dimensional space-time in general relativity negates this generalization as the lines of force would not emanate radially from the source as they do in a flat space. Since gravity is the only universal force, it would be nice to make a case that general relativity implies that space is 3-dimensional. To my knowledge, Tangherlini [10] appears to be the only one to come to grips with the dimensionality of space problem in the context of curved space-time. He concludes: "In conclusion, the above remarks suggest that it is logically advantageous to regard the dimensionality of space as a specificity to be derived from physical principles rather than simply inserted in the theory from the beginning. With further work, we may come to regard n = 3 as an eigenvalue." His gravitational arguments are quite general and complex, but do not appear compelling to Callender. Nevertheless Tangherlini is to be commended for tackling the difficult problem of the dimensionality of non-Euclidean space with such generality.

## 7  Conclusion

The analysis herein concludes that atoms (and planetary systems) as we know them, i.e. obeying known laws, cannot exist in higher dimensional space because they are not energetically bound i.e. $E_n \geq 0$. This is why higher dimensions are outside our realm of observation. The same energy equation also implies that atom-like bodies cannot exist in one and two-dimensional spaces.

A testable prediction is made regarding the search for deviations from $r^{-2}$ of the gravitational force at sub-millimeter distances. The results here imply that such a deviation must occur at $< \sim 10^{-10}$m, since atoms would fly apart if the curled up dimensions of string theory were larger than this. Considering muonium with a muon of charge $e$ orbiting a proton (for a short time before the muon decays), further reduces the limit for deviation. Since a muon is 206 times more massive than an electron, by Eq. (25) its orbital radius is 206 times smaller than an electron's orbit. Thus a deviation from $r^{-2}$ must occur at $< \sim 10^{-12}$m.

The analysis in this paper showing that atoms cannot be energetically bound for $n \geq 4$ is much stronger than the instability arguments of others. Furthermore, no testable predictions appear to be made in their papers. Callender [3] in examining scores of referenced papers says, "In general I argue that modern 'proofs' of the dimensionality of space have gone off track." By their criteria, orbits that precess (such as that of Mercury, and to a lesser extent other planets) would have to be considered unstable. Yet these orbits persist!

**Acknowledgments.** I wish to express my appreciation to Ned Britt, Felipe Garcia, Michael Ibison, Stuart Mirell, and Frank Rahn for their encouragement and helpful comments.